\documentclass{webofc}

\usepackage{url}
\hypersetup{colorlinks=true,citecolor=blue,urlcolor=blue,linkcolor=blue}

\usepackage{graphicx}
\usepackage{amsmath}
\usepackage{amssymb}
\usepackage[varg]{txfonts}   

\begin{document}

\title{The MexNICA Collaboration in the MPD-NICA Experiment at JINR: Experimental and Theoretical Achievements}

\author{Alfredo Raya\inst{1}\thanks{Corresponding author: alfredo.raya@umich.mx}, Mauricio Alvarado\inst{2}, Juan Anz\'urez\inst{1},  Alejandro Ayala\inst{2}, Wolfgang Bietenholz\inst{2}, Salom\'on Borjas Garc\'\i a\inst{3}, Eleazar Cuautle\inst{2}, Pedro E. García Gonz\'alez\inst{3}, Irving Iván Gaspar Gregorio\inst{4}, Isabel Dom\'\i nguez\inst{5}, Luis Alberto Hern\'andez\inst{4}, Maribel Herrera\inst{1,6,7}, Israel Luna\inst{2},  Pablo Mart\'\i nez-Torres\inst{3}, Emanuel Nolasco G\'omez\inst{1}, Miguel Enrique Pati\~no\inst{2},  Manuel El\'\i as Pech Dzul\inst{1}, Juan Carlos Ramírez Márquez\inst{4}, Mauricio Reyes Guti\'errez\inst{1}, Ulises S\'aenz-Trujillo\inst{1}, Roberto Tapia S\'anchez\inst{1}, Mar\'\i a Elena Tejeda-Yeomans\inst{6}, Galileo Tinoco-Santillán\inst{1}, Carlos Rafael Vázquez Villamar\inst{2} 
}
\institute{
Facultad de Ingeniería Eléctrica, Universidad Michoacana de San Nicolás de Hidalgo,\\
Edificio ``Omega 1'' Primer Piso, Ciudad Universitaria.
Av. Francisco J. Mújica S/N, C.P. 58030, Morelia Michoacán, Mexico
\and
Instituto de Ciencias
Nucleares, Universidad Nacional Aut\'onoma de M\'exico,\\ Apartado
Postal 70-543, CdMx 04510,
Mexico
\and
Instituto de Física y Matemáticas, Universidad Michoacana de San Nicolás de Hidalgo,\\
Av. Francisco J. Mújica S/N, Ciudad Universitaria, Edificio C-3, Morelia Michoacán,\\ C.P. 58040, Mexico
\and
Departamento de Física, Universidad Autónoma Metropolitana, \\
Av. San Rafael Atlixco 186, CdMx C.P. 09340, Mexico
\and
Facultad de Ciencias Físico-Matemáticas, Universidad Autónoma de Sinaloa, \\
Avenida de las Americas y Boulevard Universitarios, Ciudad Universitaria, Culiacán, 80000, Mexico
\and
Facultad de Ciencias-CUICBAS, Universidad de Colima, \\
Bernal Díaz del Castillo No. 340, Col. Villas San Sebastián, 28045 Colima, Mexico
\and Facultad de Ingeniería, Universidad Veracruzana,\\ Km. 1 Carretera Sumidero - Dos Ríos, 94458, Ixtaczoquitlán, Veracruz, Mexico
}

\abstract{
The MexNICA Collaboration coordinates the activities of Mexican scientists, engineers, postdoctoral fellows and students in the Multi-Purpose Detector experiment at the Nuclotron-based Ion Collider fAcility of the Joint Institute for Nuclear Research in Dubna, Russia. Established in 2016, the collaboration brings together five Mexican institutions whose contributions span detector development as well phenomenological and theoretical studies, including modeling by means of Monte Carlo simulations. This work summarizes the main achievements of MexNICA, consisting of the development of the miniBeBe trigger detector as well of results of phenomenological investigations of the baryon-rich region in the QCD phase diagram accessible at NICA energies, and theoretical advances based on lattice QCD and effective models. 
}
\maketitle

\section{Introduction}

The exploration of  the properties of strongly interacting matter under extreme conditions of temperature and baryon density constitutes one of the fundamental challenges in contemporary nuclear and particle physics. The phase structure of Quantum Chromodynamics remains to a large extent uncharted, particularly in the high baryon density region where lattice QCD calculations face the notorious sign problem. Understanding this region is crucial not only for fundamental physics but also for describing the interior of compact astrophysical objects  and for reconstructing the conditions of the early Universe microseconds after the Big Bang.

The MexNICA Collaboration was established in 2016 as a coordinated effort by Mexican institutions to participate in the NICA scientific program~\cite{NICAfacility}, particularly in the Multi-Purpose Detector (MPD)~\cite{MPDTDR, MPD2022,  MPDRMF}. 
The collaboration brings together professors, researchers, engineers, postdoctoral fellows, and students working across three main areas. First, experimental contributions through the development of the miniBeBe trigger detector. Second, phenomenological studies providing predictions for observables at NICA energies, including meson and baryon production ratios, pion correlation functions, photon yields in magnetic fields, baryon number fluctuations near the expected Critical End Piont (CEP), and hyperon polarization excitation functions. Third, theoretical investigations using lattice QCD simulations with the O(4) non-linear $\sigma$ model to circumvent the sign problem.

The complementarity of NICA with other facilities worldwide enables a comprehensive mapping of the QCD phase diagram across different collision energy ranges. While RHIC at Brookhaven National Laboratory explores center-of-mass energies from 7.7-200~GeV through its Beam Energy Scan program, and the LHC at CERN operates at TeV energies with low baryon chemical potential, the energy range of NICA provides unique access to the maximum baryon density region, where first-order phase transitions and a CEP are theoretically expected. The organization of this contribution is as follows: In Sec.~\ref{sec2}, we discuss the general conceptual design of the miniBeBe detector. Section~\ref{sec:pheno} summarizes the phenomenological lines of research carried out by the collaboration, whereas Sec.~\ref{sec:th} presents the theoretical efforts aligned with the scientific program of MPD. Concluding remarks are presented at the end, in Sec.~\ref{sec:conclu}.

\section{Experimental Contribution: The miniBeBe Detector}\label{sec2}

A critical requirement for the MPD experiment is an efficient Level-0 trigger system for the Time-of-Flight (TOF) detector, particularly for low-multiplicity events encountered in peripheral heavy-ion and fixed target collisions. The nominal Fast Forward Detector (FFD) of MPD demonstrates efficiency exceeding 50\%  only for central and semi-central nucleus-nucleus collisions with multiplicities above 25 particles. To address this limitation, the MexNICA Collaboration has conceptualized and developed  a mini Beam-Beam (miniBeBe) detector, a small sized detector working as to complement the TOF trigger. A detailed description of such detector can be found in Refs.~\cite{miniBeBedesign, miniBeBemechanical}.

The miniBeBe detector is designed as a fast, low-cost, low-material-budget trigger detector surrounding the interaction point. The key design specifications target a time resolution of less than 100 ps using plastic scintillators EJ232 of dimensions 20 mm $\times$ 20 mm $\times$ 5 mm coupled to Silicon Photomultipliers of the Hammamatsu series 13. The detector comprises eight electronic modules arranged in H-shaped strips with 40 SiPMs per arm, totaling 320 SiPMs across the entire system. All materials are selected to be non-ferromagnetic to avoid interference with the MPD solenoid magnetic field, and the detector must operate reliably in the 0.5  T magnetic field environment while maintaining controlled temperature conditions.

The mechanical design version 2.0 represents significant improvements over the initial conceptual design published in 2021~\cite{miniBeBedesign}. The major innovations address both physics performance and engineering practicality considerations that emerged during the initial testing and integration planning phases.
The geometry has been optimized such that SiPMs are now positioned on the narrow sides of plastic scintillators rather than the broad faces. This configuration reduces direct radiation exposure to the photosensors while maintaining excellent light collection efficiency. The repositioning also allows for better thermal management since the SiPMs are mounted closer to the cooling infrastructure, and it simplifies the mechanical mounting system by reducing the number of cable routing paths through the detector volume.
A flange-based mounting system replaces the previous rail geometry, enabling several operational advantages. Individual modules can be replaced without disrupting other MPD subsystems, which is essential for maintenance during long data-taking periods. The design is compatible with the Inner Tracking System (ITS) installation schedule, allowing the  miniBeBe to operate during Phase I of MPD operations and then be completely removed after Phase I to accommodate the ITS installation. 

Materials selection was guided by the dual constraints of non-ferromagnetic requirements and structural integrity needs. The housing employs M55-J carbon fiber composite, which provides an excellent stiffness-to-weight ratio. 
The flanges use Aluminum alloy 5052, chosen for its combination of strength, machinability, and non-magnetic properties. Cooling plates incorporate carbon fiber composite with embedded polyamide tubes for water circulation. 
The final specifications demonstrate the success of this design approach~\cite{miniBeBemechanical}. The total length amounts to 1611 mm, with a detector radius of 156 mm. Under operational loads, maximum deformation remains below 1 mm, and von Mises stress peaks below 2M Pa. The total mass is 26.18 kg, with engineering studies indicating a potential for 40-45\% reduction  
through optimization of support structure elements and thinner wall sections.

Maintaining stable operating temperatures for the SiPMs is critical for achieving the target time resolution. SiPM gain and dark count rate both depend sensitively on temperature, with typical temperature coefficients requiring control to within a few degrees Celsius for stable operation. The thermal management system incorporates both passive and active cooling elements to address heat generation from SiPM dark current, readout electronics power dissipation, and ambient heat load from surrounding detector systems.
Water cooling forms the primary active cooling mechanism. Carbon fiber cooling plates with embedded polyamide tubes circulate water at the inlet temperature of 18$^\circ$C, with the outlet temperature reaching 21.4$^\circ$C. The temperature distribution across different zones shows excellent uniformity. 
The cooling system maintains SiPM operating temperatures below 22.6$^\circ$C, well within the requirement temperature for stable performance. 
An additional gas cooling system is planned to provide supplementary temperature reduction.

Electronics and data acquisition systems are in version 2.0, with testing and optimization ongoing. The readout electronics must process signals from 320 SiPMs with timing precision of less than 100 ps, requiring careful design of amplification stages, discrimination circuits, and time-to-digital converters. Prototype electronics modules have been tested with different radiation sources, including cosmic rays, $\alpha$ and $\gamma$ radiation from radioactive sources, achieving a resolution time of less than 200 ps in laboratory tests for the lowest energy radiation and around 50 ps for high-energy muons. The high voltage power supply for the miniBeBe detector is of own design comprising low cost, low noise and high precision. The design further allows for dynamical bias voltage for each individual SiPM corrected from feedback thermal readings of PT100 sensors located along the system. Front-end electronics is designed to collect and amplify the signals from each SiPM in the detector in FERS boards to provide this information to the data collection center of the MPD. 

Numerical simulations at the level of event reconstruction of the performance of the miniBeBe have also been considered as part of the experimental and phenomenological performance of the miniBeBe detector. The results along this direction have been of the utter most importance for the mechanical and electronic design of the detector. These studies are continuously improved to reach a stage of the integration of the miniBeBe within the complete environmental landscape of simulations of the MPD experiment.

Integration is planned to be coordinated with MPD-ITS and TOF teams through regular joint working group meetings. The assembly sequence has been defined starting with mounting of ITS-compatible flanges, positioning of the miniBeBe housing on these flanges, attachment of miniBeBe-specific flanges and locking rings, installation of electronic modules, and finally fixing of cable and pipe supports for services routing. The timeline targets deployment for fixed-target mode operations beginning in mid-2026, with full commissioning 
planned for the second half of 2026.

\section{Phenomenological Contributions}\label{sec:pheno}

The MexNICA Collaboration has developed a comprehensive phenomenological program providing predictions specifically targeted for the NICA energy range. 

\subsection{Baryon-to-Meson Transition Studies}

Thermal statistical models predict a rapid transition from baryon- to meson-dominated hadronic matter as a function of collision energy. This transition occurs at temperature around 140 MeV and baryon chemical potential around 420 MeV, corresponding to center-of-mass energy per nucleon pair approximately 8.2 GeV, which is precisely within the NICA energy range. The physical origin of this transition lies in the interplay between thermal excitation favoring lighter mesons and chemical potential effects favoring baryons, with the transition occurring when these competing effects balance.

The MexNICA group has conducted Monte Carlo feasibility studies to measure the crossing points of meson and baryon transverse momentum spectra. The methodology involves generating heavy-ion collision events using the UrQMD transport model, reconstructing particle tracks through a GEANT simulation of the MPD detector, applying realistic particle identification criteria based on energy loss in the TPC and TOF measurements, and fitting the spectra to extract crossing point parameters as functions of centrality and collision energy.
The key findings demonstrate that, as the centrality of the collision decreases,  
the crossing point systematically shifts to higher transverse momenta. This behavior reflects the evolution of the freeze-out conditions, with peripheral collisions exhibiting lower temperatures but stronger radial flow effects compared to central collisions. The observable provides access to freeze-out parameters including temperature, chemical potential, and radial flow velocity through simultaneous fits for multiple particle species. The centrality dependence reveals the interplay between thermal motion and collective expansion, with the relative importance of these two effects varying across the phase space.
These studies demonstrate that MPD can precisely characterize the baryon-to-meson transition region, providing crucial constraints on the QCD phase diagram structure~\cite{MexNICAbaryonmeson}. The experimental systematic uncertainties have been estimated, including effects from particle identification efficiency, momentum resolution, and background contamination, and all are at levels that permit meaningful physics interpretation of the crossing point measurements.

\subsection{Pion Femtoscopy and Source Structure}

Two-pion correlation functions provide direct information on the space-time structure of the pion emission source through quantum statistical Bose-Einstein correlations and final-state Coulomb effects. The correlation strength and shape encode information on the source size, shape, emission duration, and relative contributions of different production mechanisms including primary production and secondary production from resonance decays.

The MexNICA group has analyzed the evolution of correlation functions with collision energy using the UrQMD event generation combined with the CRAB correlation afterburner code that properly accounts for quantum statistics and Coulomb final-state interactions. The simulation chain includes the generation of collision events, tracking of all hadrons through their decays and rescatterings, identification of pion pairs suitable for correlation analysis based on their momenta and charges, and calculation of correlation functions as functions of the pair relative momentum and the pair average momentum.
The analysis reveals that Lévy-stable distributions provide superior descriptions compared to Gaussian or exponential shapes at NICA energies~\cite{MexNICAfemtoscopy}, indicating the presence of large tails in the emission source distribution. This behavior suggests contributions from multiple length scales in the source, consistent with a core-halo picture, where a compact central region is surrounded by an extended halo of particles from long-lived resonance decays.

Separation of pions into primary and secondary components based on their production mechanisms reveals a systematic ordering of source sizes. Secondary pions exhibit larger source radii than the inclusive sample, which in turn shows larger radii than the primary pion sample produced from short-lived resonance decays. This hierarchy confirms that different production mechanisms contribute differently to the femtoscopic observables, and a proper interpretation requires accounting for these multiple components.
The core-halo structure can be quantitatively characterized through analysis of the correlation function intercept parameter, which measures the fraction of pions coming from the core region versus the extended halo. The results indicate that core pions originate primarily from slow-moving resonance decays such as $\omega$ and $\eta$ mesons, while a smaller fraction comes from primary production processes. The evolution of the core fraction with collision energy and centrality provides information about the freeze-out dynamics and the degree of thermalization achieved in the collision.
The Lévy stability index, which characterizes the power-law tail behavior of the source distribution, has been proposed as a potential probe for critical fluctuations near the Critical End Point (CEP), where long-range correlations develop in the system, manifesting as enhanced tails in the source distribution. The MexNICA studies provide baseline predictions for the Lévy index as a function of collision energy in the absence of critical effects, against which either simulations  or experimental measurements can be compared to search for some anomalous behavior that might signal proximity of the CEP.

\subsection{Photoproduction in Magnetic Fields}

Non-central heavy-ion collisions generate intense transient magnetic fields through the charged spectator nucleons that pass by the interaction region at nearly the speed of light. At NICA energies, these fields can reach magnitudes on the order of the pion mass squared at early times, which means that $|eB|$ approaches  20,000 MeV$^2$. The fields peak during the pre-equilibrium phase when gluon occupancy is maximal, creating conditions where magnetic field effects on QCD processes can be significant.

The MexNICA group, in collaboration with researchers from other Latin American institutions, has developed a comprehensive theoretical framework for photoproduction via gluon-driven processes enhanced by the magnetic field~\cite{MexNICAphoton1}. Two key mechanisms have been identified: gluon fusion and gluon splitting processes.  
Both of these processes are absent in vacuum but become enhanced in the presence of strong magnetic fields due to modifications of the quark propagators and vertices, which ultimately originates from the loss of $C$ and $P$, as well as of Lorentz invariance.
The magnetic field induces anisotropy in the gluon pressure, which  
affects the pre-equilibrium dynamics by accelerating isotropization of the momentum distribution compared to scenarios without magnetic fields. The faster approach to local momentum isotropy has implications for the applicability of hydrodynamic descriptions at early times and affects the time evolution of photon production rates.
Current work extends these calculations to the intermediate magnetic field regime where $|eB|$ is comparable to $\Lambda_{\rm QCD}^2$, relevant for longer-lasting field configurations that persist beyond the initial pre-equilibrium phase into the early hydrodynamic evolution. The intermediate regime requires careful treatment of the interplay between magnetic field effects and the temperature-dependent QCD coupling, as both become comparable in magnitude. Predictions for differential photon yields and elliptic flow coefficients at NICA energies are under scrutiny, as they  
provide experimental targets for the MPD photon detection capabilities.

\subsection{Search for the Critical End Point}

Locating the QCD CEP remains one of the primary goals of the heavy-ion collision program worldwide. The CEP marks the end of the crossover  line in the temperature-baryon chemical potential plane, and its location determines the nature of the transition accessible experimentally. Lattice QCD calculations establish that at vanishing baryon chemical potential, the transition is a crossover occurring at pseudocritical temperature approximately 155 MeV. Model calculations using various effective theories predict that this crossover transition becomes first-order at high baryon chemical potential, thus predicting the existence of a CEP at intermediate values of temperature and baryon chemical potential.

The MexNICA group employs the Linear Sigma Model with quarks to study CEP signatures. This effective model includes scalar and pseudoscalar meson degrees of freedom coupled to constituent quarks, with parameters fit to vacuum phenomenology including pion mass, pion decay constant, and $\sigma$ meson properties. The model incorporates the essential symmetries of low-energy QCD, i.e., chiral symmetry and its spontaneous breaking, allowing for a systematic exploration of the finite temperature and density phase diagram.

Baryon number fluctuations characterized by cumulants provide sensitive probes of critical behavior. The cumulants are defined as derivatives of the pressure with respect to baryon chemical potential at fixed temperature, and near the critical point they exhibit strong divergences due to the long-range correlations developing in the system. The ratios of fourth to second cumulants and third to first cumulants are predicted to show characteristic non-monotonic behavior as functions of the collision energy, with peaks indicating proximity to the critical point. The MexNICA studies map out the expected cumulant ratio evolution across the NICA energy range for different possible locations of the CEP, providing discrimination power between different scenarios~\cite{MexNICACEP}.

The squared speed of sound $c_s^2$, defined as the derivative of pressure with respect to energy density, develops a minimum near the CEP, because the system becomes highly compressible near criticality, with large fluctuations in density occurring spontaneously. Measurement of $c_s^2$ through collective flow observables, particularly the slope of directed flow as a function of rapidity and the excitation function of elliptic flow, can provide independent constraints on the CEP location complementary to fluctuation measurements.

Vorticity effects on the phase transition have also been investigated by the MexNICA group. The intense vortical motion present in non-central collisions at NICA energies potentially affects the phase structure by coupling to the spin degrees of freedom of the medium. Calculations indicate that rotation can shift the CEP position in the temperature-baryon chemical potential plane, with the magnitude and direction of the shift depending on the strength of the vorticity and its orientation relative to the reaction plane. Understanding these effects is essential for proper interpretation of fluctuation measurements in experiments where vorticity is necessarily present.

\subsection{Hyperon Polarization Predictions}

The observation of global $\Lambda$  polarization in heavy-ion collisions at RHIC revealed that the orbital angular momentum from non-central collision geometry is partially converted into the spin polarization of produced particles. This phenomenon arises from spin-orbit coupling in the strongly interacting medium, where the thermal vorticity field generated by the collective rotation of the system aligns the spins of strange quarks, which subsequently hadronize into observable $\Lambda$  hyperons that retain memory of this initial spin alignment.

The MexNICA group developed the core-corona model to explain both $\Lambda$ and $\bar\Lambda$  global polarization in semi-central heavy-ion collisions. The model divides the collision system into a high-density core region that undergoes thermalization and collective expansion described by relativistic hydrodynamics, and a lower-density corona region dominated by hadronic rescattering described by transport models. The relative contributions from core and corona evolve with collision energy, with lower energies having larger corona fractions due to reduced stopping power and less complete thermalization.

Strange quark spin alignment with thermal vorticity proceeds through spin-orbit interactions encoded in the quark self-energy at finite temperature and density. The alignment rate depends on the relaxation time that characterizes how quickly the strange quark spin adjusts to the local vorticity field direction. The MexNICA group has calculated this relaxation time using field-theoretical methods, finding that it decreases with increasing temperature and baryon density, meaning that alignment becomes more efficient at higher densities and temperatures, where the coupling between spin and vorticity strengthens.
The volume and lifetime of the quark-gluon plasma phase are modeled using Bjorken expansion, which describes the longitudinal expansion of the system under the assumption of boost invariance. The thermal vorticity field in this scenario arises from gradients in the transverse flow velocity and depends on the collision centrality through the initial angular momentum and impact parameter. More peripheral collisions generate stronger vorticity due to larger angular momentum per participant nucleon, but they also have smaller volumes and shorter lifetimes for the thermalized core region.
The key prediction from this framework is that $\Lambda$ and $\bar \Lambda$ global polarization should reach maximum values at collision energies accessible to NICA and HADES~\cite{MexNICApolarization1}.
This peak arises from the competition between an increasing vorticity magnitude at lower energies, due to more stopping, and a larger impact parameter asymmetry, decreasing the formation time allowing more efficient spin-vorticity coupling before hadronization, and evolving core-corona relative abundances that affect the fraction of $\Lambda$s retaining the spin alignment signal from the core region.
The predicted peak structure is unique to the NICA energy range and represents a flagship measurement for the MexNICA physics program. Recent data from HADES show hints of increasing polarization toward lower energies, consistent with the predicted rising edge of the peak, while STAR data at higher RHIC energies show the falling edge. The NICA energy range covers precisely the peak region, offering the opportunity to map out the maximum polarization and thereby constraining the vorticity magnitude and the efficiency of spin-vorticity coupling under conditions of maximum baryon density.

\subsection{Microscopic Vorticity-Spin Transfer Mechanisms}

Understanding the microscopic mechanism by which macroscopic vorticity transfers to individual quark spins requires field-theoretical calculations at finite temperature and density that go beyond the hydrodynamic description. The MexNICA group has computed the relaxation time that a strange quark takes to align its spin with the direction of angular momentum in a rotating medium~\cite{MexNICAvorticity1, MexNICAvorticity}, using the real-time formalism for thermal field theory extended to include non-inertial reference frames.
The calculation involves evaluating the strange quark self-energy including thermal loop corrections modified by the presence of rotation. The imaginary part of the self-energy determines the relaxation rate through which the spin distribution approaches the equilibrium distribution aligned with the vorticity. This rate depends on temperature through the thermal occupation numbers of the interacting gluons and quarks, on baryon chemical potential through modifications of the quark dispersion relations and Fermi blocking effects, and on the rotation rate itself through centrifugal and Coriolis corrections to the propagators.
The results show that the relaxation time decreases with increasing temperature and baryon density, indicating that spin alignment becomes more efficient in denser and hotter regions of the collision. This finding has important implications for understanding which regions of the space-time evolution contribute most significantly to the observed hyperon polarization. The shorter relaxation times at high density mean that the early high-density phase contributes more strongly to spin alignment, even though it occupies a shorter time duration, while the later lower-density phase has longer relaxation times and thus contributes less despite its longer duration.
Applications of these microscopic calculations include obtaining better quantitative estimates of polarization magnitudes that can be compared directly with experimental measurements, understanding the temperature and density dependence of spin transfer efficiency to guide optimization of experimental collision energy scans for maximum signal, connecting the measured polarization to the actual angular momentum distributions present in the collision geometry, and refining parameters in the core-corona model to achieve consistency between hydrodynamic and transport descriptions of the spin dynamics.

\section{Theoretical Contributions}\label{sec:th}

Direct lattice QCD calculations at finite baryon density face the notorious sign problem, where the fermion determinant becomes complex-valued, rendering standard Monte Carlo importance sampling ineffective. The severity of this problem increases exponentially with volume and chemical potential, making it impossible to simulate realistic system sizes at NICA-relevant baryon densities using standard lattice QCD methods. Alternative approaches are therefore essential for making first-principles predictions about the phase structure at finite baryon density.

The MexNICA collaboration employs the $O(4)$ non-linear $\sigma$ model as an effective theory for two-flavor QCD in the chiral limit. This model exhibits the same spontaneous symmetry breaking pattern as QCD, specifically the breaking of $O(4)$ symmetry to $O(3)$ in the effective theory corresponds to the breaking of chiral symmetry from $SU(2)_L\times SU(2)_R$  to the diagonal $SU(2)_V$ in QCD. This correspondence places the two theories in the same universality class near the phase transition, meaning that they share the same critical behavior including critical exponents and scaling functions.
At high temperature, dimensional reduction applies, allowing the four-dimensional QCD theory to be approximated by the three-dimensional $O(4)$ model with appropriate matching of parameters. Crucially for the sign problem, topological charge in the $O(4)$ model can serve as a proxy for baryon number following the original observation of Skyrme that topological solitons in meson theories carry baryon quantum numbers. This identification allows the introduction of baryon chemical potential as a coupling to the topological charge density, and remarkably this formulation involves only real action terms, completely avoiding the sign problem that plagues standard lattice QCD at finite density.
Monte Carlo simulations using standard algorithms can therefore explore the temperature-baryon chemical potential phase diagram without encountering the exponential signal-to-noise degradation that afflicts direct QCD simulations. The MexNICA group has performed extensive simulations mapping out the critical line, corresponding to a chiral 2-flavor QCD~\cite{MexNICAlattice, MexNICAlattice2}, defined as the locus of points in the temperature-baryon chemical potential plane where the correlation length diverges, signaling either a second-order phase transition, expected to end in the CEP and turning into a  first-order phase transition line.
The results show that critical temperature decreases monotonically with increasing baryon chemical potential across the explored range extending up to approximately 309 MeV. At this maximum accessible chemical potential, the critical temperature remains above approximately 106 MeV, indicating that the chiral restoration transition remains at relatively high temperature even at substantial baryon density. No clear signature of a CEP was found within the explored region, meaning that if the CEP exists, it likely lies at either a higher baryon chemical potential beyond the currently accessible range or at a lower temperature where finite-volume effects become more severe and require larger lattices for reliable extrapolation.
Hints from the simulations suggest that the CEP may lie near the boundary of the accessible domain, specifically near the maximum explored chemical potential. The hint for a first-order transition to be near-by the highest
baryon density that we could explore is based on the
quasi-discontinuous behavior of quantities, which are given
by first derivatives of the free energy.
Extended simulations at higher chemical potential are planned but require larger computational resources due to increased correlation times and finite-volume effects.
This approach provides non-perturbative constraints on the phase diagram structure despite the sign problem, complementing effective model studies and providing benchmarks for validating the predictions from models like the Linear Sigma Model with quarks employed in the phenomenological studies described earlier. The combination of multiple theoretical approaches, each with different systematic uncertainties, provides more robust predictions than any single method alone.

\section{Conclusions}\label{sec:conclu}

The MexNICA Collaboration represents a comprehensive Mexican initiative contributing to the forefront of nuclear physics research through multifaceted participation in the Multi-Purpose Detector experiment at the Nuclotron-based Ion Collider fAcility at the Joint Institute for Nuclear Research. The collaboration has established itself across three interconnected pillars that together provide a complete program from fundamental theory through phenomenological predictions, computational simulations and experimental implementation.

Mexico's participation in the MPD–NICA program through the MexNICA collaboration represents a strategic investment in large-scale international scientific infrastructure. By contributing to detector development, Mexican institutions are acquiring advanced expertise in fast timing detectors, SiPM readout electronics, mechanical design, and integration in high-radiation and strong magnetic-field environments, capabilities that are transferable to other science and technology initiatives. The collaboration has strengthened Mexico’s high-performance computing capacity through large-scale Monte-Carlo simulations, lattice-QCD calculations, and distributed data-analysis pipelines aligned with international standards with shared computing resources amongst international partnerships. 

MexNICA has also promoted and facilitated the co-signing of Memoranda of Understanding between SECIHTI and JINR, expanding cooperation beyond nuclear physics into broader scientific and medical fields, including instrumentation, accelerator science, and training programs. Sustained student and postdoctoral mobility to JINR provides hands-on training in accelerator-based science and fosters long-term international research networks for early-career scientists. These efforts enhance the national visibility of Mexican science through publications in leading journals, contributions to MPD technical design reports and white papers, and leadership roles within international working groups. Collectively, MexNICA positions Mexico as an active partner in frontier research on QCD matter under extreme conditions while building human capital, technological capacity, and international scientific reputation that extends well beyond nuclear physics.

The experimental pillar centers on the miniBeBe detector development, which showcases Mexican engineering and technical expertise in producing a crucial trigger system for low-multiplicity events that extends the MPD physics reach to peripheral collisions and lighter collision systems. The mechanical design version 2.0 represents a mature solution validated through rigorous Finite Element Method analysis, meeting all stringent requirements for operation in the MPD environment including non-ferromagnetic material constraints, low material budget to minimize particle trajectory perturbations, efficient thermal management to maintain stable SiPM operation, and structural robustness to withstand operational loads and handling stresses. The design maintains compatibility with future detector upgrades by enabling complete removal after Phase I to accommodate ITS installation, demonstrating forward thinking that considers the long-term MPD evolution. Numerical simulations at the level of event reconstruction carried out by the MexNICA Collaboration reinforce the expected performance of the detector.

The phenomenological pillar provides targeted predictions for the baryon-rich region of the QCD phase diagram accessible at NICA energies. Studies of baryon-to-meson transition identify specific observables including the crossing point of transverse momentum spectra that can quantify the freeze-out properties and locate the transition region in collision energy. Pion femtoscopy analysis using Lévy-stable distributions reveals the source structure and provides tools for searching for critical fluctuations that would signal proximity to the CEP. Photoproduction predictions in strong magnetic fields open a new window on pre-equilibrium dynamics and gluon-dominated processes during the earliest collision stages. CEP searches using cumulant ratios and speed of sound signatures provide multiple independent observables that together can triangulate the critical point location. Hyperon polarization predictions establishing a peak at NICA energies represent a flagship measurement uniquely accessible in this energy range, with the core-corona model providing a theoretical framework connecting the observed polarization to the vorticity structure and freeze-out dynamics. These  studies, often conducted in collaboration with Latin American partners, position MexNICA as a leading contributor to theoretical understanding of baryon-rich QCD matter.

The theoretical pillar employs non-perturbative approaches to explore the fundamental structure of the QCD phase diagram and related questions in field theory. Lattice QCD simulations using the $O(4)$ model circumvent the sign problem,  
enabling first-principles exploration of the critical line as a function of baryon chemical potential and providing constraints on possible CEP locations.

As the MPD approaches commissioning for fixed-target mode in mid-2026 followed by collider mode, the MexNICA Collaboration stands ready to contribute meaningfully to unraveling the mysteries of strongly interacting matter under extreme conditions. The phenomena studied at NICA are relevant to diverse areas including the early Universe conditions, neutron star interior, and the fundamental structure of the QCD vacuum. 

The  multi-faceted approach of MexNICA, combining experimental innovation with phenomenological prediction and theoretical insight, positions Mexican science at the forefront of this exciting field. The sustained commitment from five Mexican institutions, the support from national funding agencies including CONAHCYT/SECIHTI, the partnerships with JINR and international collaborators, and the enthusiasm of students and early-career researchers all contribute to making MexNICA a successful model for Mexican participation in large-scale international scientific projects. The experience and expertise developed through MexNICA will benefit the broader Mexican scientific community and can be applied to future opportunities in particle physics, nuclear physics, and related fields requiring similar combinations of detector technology, large-scale computing, and theoretical sophistication.

\section*{Acknowledgements}
We acknowledge the MPD Collaboration for welcoming MexNICA contributions and for productive technical collaborations on detector integration and physics analysis planning. Some of the members of MexNICA acknowledge support from CONAHCyT/SECIHTI (Mexico) under grants  CF-2023-G-433, CBF-2025-G-1718 and CIORGANISMOS-2025-17. AA also acknowledges support under grant  PAPIIT-DGAPA IG100826. AR acknowledges support from CIC-UMSNH under grant 18371.

\end{document}